\documentclass[prl,twocolumn,amsmath,amssymb,showpacs,floatfix,nofootinbib,preprintnumbers,superscriptaddress]{revtex4-1}
\usepackage[latin1]{inputenc}
\usepackage{amsmath}
\usepackage{amssymb}
\usepackage{graphicx}
\usepackage{dcolumn}
\usepackage{subcaption}
\usepackage{bm}
\usepackage{latexsym}
\usepackage{epsfig}
\usepackage[normalem]{ulem}
\usepackage{graphicx}
\usepackage[colorlinks=true]{hyperref}
\usepackage[dvipsnames]{xcolor}

\renewcommand{\vec}[1]{\bm{#1}}

\begin{document}
\title{Beyond the classical distance-redshift test: cross-correlating redshift-free standard candles and sirens with redshift surveys}
\author{Suvodip Mukherjee}\email{smukherjee@flatironinstitute.org}
\affiliation{Center for Computational Astrophysics, Flatiron Institute, 162 5th Avenue, New York, NY 10010, USA}
\affiliation{Institut d'Astrophysique de Paris (IAP), UMR 7095, CNRS/UPMC Universit\'e Paris 6, Sorbonne Universit\'es, 98 bis boulevard Arago, F-75014 Paris, France}
\affiliation{ Institut Lagrange de Paris (ILP), Sorbonne Universit\'es, 98 bis Boulevard Arago, 75014 Paris, France}
\author{Benjamin D. Wandelt}\email{bwandelt@flatironinstitute.org}
\affiliation{Center for Computational Astrophysics, Flatiron Institute, 162 5th Avenue, New York, NY 10010, USA}
\affiliation{Institut d'Astrophysique de Paris (IAP), UMR 7095, CNRS/UPMC Universit\'e Paris 6, Sorbonne Universit\'es, 98 bis boulevard Arago, F-75014 Paris, France}
\affiliation{ Institut Lagrange de Paris (ILP), Sorbonne Universit\'es, 98 bis Boulevard Arago, 75014 Paris, France}
\affiliation{Department of Astrophysical Sciences, Princeton University, Princeton, NJ 08540, USA}
\date{\today}
\begin{abstract}
LSST will supply up to $10^6$ supernovae (SNe) to constrain dark energy through the distance--redshift ($D_L$--$z$) test. Obtaining spectroscopic SN redshifts (spec-$z$s) is unfeasible; alternatives are suboptimal and may be biased. We propose a powerful multi-tracer generalization of the Alcock-Paczynski test that pairs redshift-free distance tracers and an overlapping galaxy redshift survey. Cross-correlating $5\times 10^4$ redshift-free SNe with DESI or Euclid outperforms the classical $D_L$--$z$ test  with spec-$z$s for all SN. Our method also applies to gravitational wave sirens or any redshift-free distance tracer.
\end{abstract}
\maketitle
Since the discovery of cosmic acceleration using type Ia supernovae (SNe) \cite{Perlmutter:1998np, 1998AJ....116.1009R, Riess:2004nr}  as standard candles, the luminosity distance--redshift ($D_L$--$z$) relation has stood as a pillar supporting the  Lambda Cold Dark Matter (LCDM) cosmology, together with the Cosmic Microwave Background (CMB) and other probes of cosmic Large Scale Structure (LSS).  And yet, an understanding of the  physics of cosmic acceleration, e.g. in terms of a hypothesized dark energy component or modifications of general relativity, remains elusive. An accurate trace of the expansion history, e.g. through the $D_L$--$z$ relation, is one of the foremost goals of current and next-generation surveys such as SDSS-IV \cite{sdss-iv}, DES \cite{Abbott:2005bi}, DESI \cite{Levi:2013gra, Aghamousa:2016zmz}, EUCLID  \cite{Amiaux:2012bt}, LSST \cite{Abell:2009aa,Marshall:2017wph}, and WFIRST \cite{Spergel:2015sza}.

LSST is going to generate a large type Ia SN sample at a rate of $\ge 10^4\, \text{yr}^{-1}$,  an order of magnitude greater than the total number of SNe currently known. With this exponential increase in data, SN cosmology must move into a new regime. Already now, the $\ge10^3\, \text{yr}^{-1}$ SNe that are being observed over a wide redshift range make it impossible to  obtain time-consuming  spectroscopic follow-up for every SN, the traditional approach underlying the success of cosmology with standard candles over the last two decades  \cite{Scolnic:2017caz}.
For the majority of SNe in current and upcoming  surveys, type and redshift must be identified using photometry alone.  An alternative approach combines photometric types with spectroscopic redshift measurements of the presumed SN host galaxy\cite{Gupta:2016ods}. Errors may lead to biases and  loss of information in the inferred cosmological parameters.

At the same time, future galaxy redshift surveys like  DESI \cite{sdss-iv} and EUCLID \cite{Amiaux:2012bt} are going to measure tens of millions of galaxy redshifts over large fractions of the sky. As a result, it is realistic to expect a galaxy catalog with $\sim 10^7$ spectroscopic redshifts overlapping SN data sets over a wide redshift range on the time scale of LSST. 

In this Letter, we  propose a new method to  infer  cosmological parameters  accurately from distance tracers (e.g. SNe) \textit{without} redshifts. The main idea is to exploit the fact that both distance tracers  and galaxies are tracers of the matter density, and therefore spatially correlated through the underlying matter field.  Aided by the reduced shot noise in upcoming large SN samples we can therefore tightly constrain cosmology by maximizing the spatial cross-correlation of overlapping distance catalog and redshift catalogs. More broadly, our approach shows a classical cosmological test in a new light as a limit of a multi-tracer generalization of the Alcock-Paczynski (A-P) test \cite{1979Natur.281..358A}.

A particular feature of this cross-correlation approach is its robustness to both data systematics and modeling assumptions. It is insensitive to SN photometric redshift (photo-$z$) errors by construction. It is also more robust to type contamination and  opens up new ways to assess the impact of astrophysical systematics on SN cosmology.  Theoretical error in the models of the isotropic galaxy and SN auto/cross-correlations  does not lead to estimator bias but will affect the optimality of the estimate; in addition the approach will be less sensitive  to approximations  such as neglecting non-linear redshift space distortions (RSDs).
In contrast to $dN/dz$ determinations based on clustering redshifts \cite{Menard:2013aaa}, a technique to infer cosmological $z$ distributions $dN/dz$ when one tracer has very poor distance information (see also\cite{Newman:2008mb, Schmidt:2013sba, Rahman:2015ecl, Scottez:2016fju, Scottez:2017glm}), in our case both galaxies and SNe have  radial measurements (redshifts and distances, respectively). This avoids a potential failure mode when clustering bias varies significantly over the scale of the radial uncertainties. 

\paragraph{Setup.} Type Ia SNe discovered in cosmological surveys constrain the $D_L$ through the observed apparent magnitude ($\hat m$) and the absolute magnitude (M), calibrated from photometric observations of the SN lightcurve, through the relation
\begin{equation}\label{mag-dl}
m= 5\log_{10}\bigg(\frac{D_L(z)}{\text{pc}}\bigg) +M -5.\\
\end{equation}
 $D_L(z)$ is related to the cosmological model and to the redshift $z$ by 
\begin{equation}
\begin{split}
D_L(z)&=\frac{c}{H_0}(1+z)\int_0^z\frac{dz'}{\sqrt{\mathcal{E}(z)}}\label{eqdl},
\end{split}
\end{equation}
where $\mathcal{E}(z) \equiv \Omega_m(1+z')^3 + \Omega_{de}e^{3\int_0^z d\ln(1+z')(1+w(z'))}$. To illustrate our approach we here assume a flat ($\Omega_K = 0$, i.e $\Omega_{de}= 1- \Omega_{m}$) Universe with dark energy equation of state as  $w(z) = w_0 + w_a(z/(1+z))$; but what follows applies to  scenarios with other  models of cosmic acceleration and non-zero curvature. 

The isotropic two-point correlation function at a separation $r$ between two tracers ($x$ and $y$) of the density fluctuations, {e.g., $1+\delta_{x} (\bm{s})= \rho_{x}(\bm{s})/\bar \rho$, with respect to the background density $\bar \rho$ can be written as}
{\begin{equation}\label{corr-gal}
\xi^{iso}_\textrm{x-y}(r)= \frac{1}{2\pi^2} \int k^2 dk\, b_x (z) \, b_y (z)\,  P(k)j_0(kr) e^{-k^2/k_{max}^2},
\end{equation}}
where, $P(k)$ is the non-linear power spectrum {obtained from the ensemble average} of the density fluctuations  in the Fourier domain for wavenumber $\bm{k}$, $j_0(kr)$ is the spherical Bessel function and $b_x= \delta_x/\delta_{dm}$ is the bias of  tracer $x$ with respect to dark matter. The cutoff $k_{max}$ is introduced for a numerical convergence at high $k$, to avoid the oscillatory behavior of $j_0(kr)$ \cite{2012MNRAS.427.3435A}.
For galaxies, $x={g}$, with galaxy bias $b_ {g}\approx 1.6$, whereas for  SNe, $x={sn}$ and $b_{{sn}}= \delta_{{sn}}/\delta_{dm}$. While $b_{{sn}}$  is uncertain there are studies which indicate that the $b_{{sn}}$ may exceed $b_ {{g}}$ by around $60\%$ \cite{Carlberg:2008qf}; we will conservatively take  $b_{{sn}}\approx 1.6$  as a fiducial value. Note that the specific value of $b_{{sn}}$ only affects the precision of parameter inferences with higher bias giving tighter constraints; mis-specifying it will not cause error.

RSDs will affect galaxy positions in redshift space   \cite{1987MNRAS.227....1K, 1992ApJ...385L...5H, PhysRevD.70.083007}. In our approach the SNe do not carry any redshift label, so RSDs do not affect them (in contrast to the traditional $d_L$--$z$ test). The galaxy-SN cross-power spectrum therefore has just a single Kaiser factor \cite{1987MNRAS.227....1K}, $P_{gs}(k)= (1+ f\mu^2_k/b_{g}) P(k)$.

Here, $f\equiv d\ln G/d\ln a$ is the dimensionless growth rate, $G$ is the growth factor, and $\mu_k$  is the cosine of the angle between the Fourier modes and the line of sight. The anisotropic cross-correlation function  is \cite{1992ApJ...385L...5H}
\begin{align}\label{corr--rsd}
\begin{split}
\xi_{\textrm{g-sn}}(\vec r)= &\bigg(1+ \frac{f}{3b_{g}}\bigg)\xi^{iso}_\textrm{g-sn}(r) \mathcal{P}_0(\mu_r)  \\& + \frac{2f}{3b_{g}}(\xi^{iso}_\textrm{g-sn}(r)- \bar \xi_{gs}( r)) \mathcal{P}_2(\mu_r), \\ &\text{where}\, 
\bar \xi_\textrm{g-sn}(r)=  \frac{3}{r^3}\int_0^r \xi^{iso}_\textrm{g-sn}(s) s^2 ds,
\end{split}
\end{align}
and $\mu_r$ is the cosine of the angle between the line of sight and $\vec r$ and $\mathcal{P}_\ell(\mu_r)$ is the $\ell^\text{th}$ order Legendre polynomial. The anisotropic $\textrm{g-g}$ and $\textrm{sn-sn}$ autocorrelations take analogous forms.

\paragraph{Method.}\label{algorithm}
Consider both galaxies and SNe as  tracers of $\delta_{dm}$ in comoving coordinates. Galaxies are observed in redshift space and SNe in the $D_L$ space. Modeling these observed galaxy and SN overdensities $\bm{\delta}_{g,sn}$ as correlated Gaussian random fields, the log-likelihood for $\bm{\theta} \equiv \{\Omega_m, w_0, w_a, H_0\}$ becomes
\begin{equation}
-2\mathcal{L}_\text{full}(\bm{\delta}_g,\bm{\delta}_{sn}|\bm{\theta})={\begin{pmatrix} 
	\bm{\delta}_g  \\
    \bm{\delta}_{sn}
\end{pmatrix}^T
\bm{\Xi}^{-1}
\begin{pmatrix}
	\bm{\delta}_g\\
	\bm{\delta}_{sn}
\end{pmatrix}
+\ln{|\bm{\Xi}|}}.\label{eq:fullloglike}
\end{equation}
The covariance matrix $\bm{\Xi}(\bm{\theta})$  takes the block matrix form
\begin{equation}
\bm{\Xi}(\bm{\theta})=\begin{pmatrix}
        \bm{Z}^T(\bm{\theta})\bm{\xi}_\textrm{g-g}\bm{Z}(\bm{\theta}) &
          \bm{Z}^T(\bm{\theta})\bm{\xi}_\textrm{g-sn}\bm{D}(\bm{\theta})\\
        \bm{D}^T(\bm{\theta})\bm{\xi}_\textrm{g-sn}^T\bm{Z}(\bm{\theta}) &
          \bm{D}^T(\bm{\theta})\bm{\xi}_\textrm{sn-sn}\bm{D}(\bm{\theta})
\end{pmatrix},
\end{equation}
where the components of the $\bm{\xi}_\textrm{x-y}$ are computed from equations analogous to
Eq.~(\ref{corr--rsd}).
The  parameter dependence in this likelihood enters exclusively in $\bm{\Xi}$ through the mappings $\bm{Z}$ and $\bm{D}$ from comoving coordinates to $z$ and $D_L$, respectively. The form of the $\bm{\xi}$ are assumed fixed to  a fiducial cosmology; the optimal parameters will choose mappings $\bm{Z}$ and $\bm{D}$ that best remap the (isotropic, except for RSD) forms of the correlation functions into the data space. 
Consequently, the $\textrm{g-g}$ block corresponds to the standard galaxy auto-correlation A-P test\cite{2016ApJ...832..103L}, while the $\textrm{sn-sn}$ block is a generalized A-P test based on SN luminosity distances instead of galaxy redshifts. The standard A-P test is known to be robust to  the details of $\xi$ \cite{Kodi-RamanahInPrep}; we expect this to carry over to this more general version. The joint analysis including auto and cross terms will lead to sample variance cancellation (owing to the same underlying $\delta_{dm}$ being probed by both tracers). We will explore these features of our approach in future work. 

For clarity of presentation, the remainder of this letter will focus  on a simplified treatment of the cross-correlation A-P constraints on the $D_L$--$z$ relation. Rather than work with Eq.~(\ref{eq:fullloglike}) we will directly construct a likelihood for the  $\textrm{g-sn}$ correlation to forecast parameter constraints. For any $\bm{\theta}$ we can remap both galaxies and SNe into comoving coordinates $\bm{r}_i$ and estimate $\hat \xi_\textrm{g-sn}(\bm{r}_{ij})(\bm{\theta})$.  The likelihood will peak for parameters that best remap the SNe and the galaxies in such a way that their spatial structures match each other. Near the peak we can write an approximate, Gaussian log-likelihood 
\begin{equation}
-2\mathcal{L}_\textrm{g-sn}=\sum_{\text{all}\,i,j}({\hat \xi_\textrm{g-sn} ( \vec{r}_{ij}(\bm{\theta}))}-{\xi^\ast_\textrm{g-sn}}(\vec{r}_{i'j'}))C^{-1}_{iji'j'}(ij\rightarrow i'j').
\end{equation}
${C}$ is a model of the covariance of the correlation function (cf Eq.~\eqref{errormodel-1} below). 

To forecast parameter constraints, we calculate the Fisher matrix  for this log-likelihood \cite{Tegmark:1996bz,1997PhRvD..55.5895T, Heavens:1999am, Alsing:2017var}
\begin{align}\label{fisher-error-1}
\begin{split}
\bm{F}_{\theta_a\theta_b}=& \sum_{\text{all}\,i,j} \partial_{\theta_a}\xi_\textrm{g-sn} ( \vec{r}_{ij}(\bm{\theta}))\vert_{\theta_{fid}}
\bm{C}^{-1}\partial_{\theta_b}\xi_\textrm{g-sn} ( \vec{r}_{ij}(\bm{\theta}))\vert_{\theta_{fid}}
\end{split}
\end{align}
using the chain rule.

\paragraph{Parameter Forecasts.} \label{results}
Using the above prescription, we forecast the parameter constraints for an upcoming LSST-like SN catalog. Awaiting a detailed simulation study to assess the effects of off-diagonal terms in the covariance matrix we approximate the dominant term of the covariance matrix as diagonal with elements \cite{1993ApJ...412...64L} 
\begin{equation}\label{errormodel-1}
\begin{split}
C(\vec r,\vec{r})= \frac{(1+\xi_\textrm{g-sn}(\vec r))^2}{N^\textrm{sn}_p}, 
\end{split}
\end{equation}
where 
$N^\textrm{sn}_p$ 
is the number of pairs of SN-galaxy samples. 

The  $D_L$ error $\sigma_{D_L}$ obtains from the intrinsic scatter $\sigma_{\text{int}}$ by the magnitude-distance relation given in Eq. \eqref{mag-dl} \cite{Scolnic:2017caz}). We model its effect on $\delta_\textrm{sn}$ as an anisotropic smearing along the line of sight in the Fourier domain:  $\delta_ke^{-k^2\sigma^2_\parallel \mu_k^2/2}$, where $\mu_k\equiv \hat k.\hat n$ and  $\sigma_\parallel\equiv \sigma_{D_L}/(1+z)$.
In this analysis we have taken a constant $\sigma^2_{\text{int}} = 0.12$ mag according to LSST forecasts \cite{Abell:2009aa,Marshall:2017wph} and the correlation function is calculated using Eq. \eqref{corr--rsd} with the non-linear power spectrum from CLASS \cite{2011arXiv1104.2932L, 2011JCAP...07..034B, 2011JCAP...10..037A}. 

We have assumed that the  galaxy redshifts are spectroscopic, leading to negligible error $\sigma_z =10^{-4}$. We introduce a k-space cut-off $k_{max}=1h$ Mpc$^{-1}$ in Eq.~\eqref{corr-gal}; we discuss in the conclusions why this choice is likely to be conservative. To assess the possible impact of stochastic small-scale effects we also show the results for a even more stringent cut-off $k_{max}=0.5h$ Mpc$^{-1}$.

\begin{figure}[t]
\centering
\begin{subfigure}{1.05\linewidth}
\includegraphics[trim={0 0 0 0cm}, clip, width=1.\linewidth]{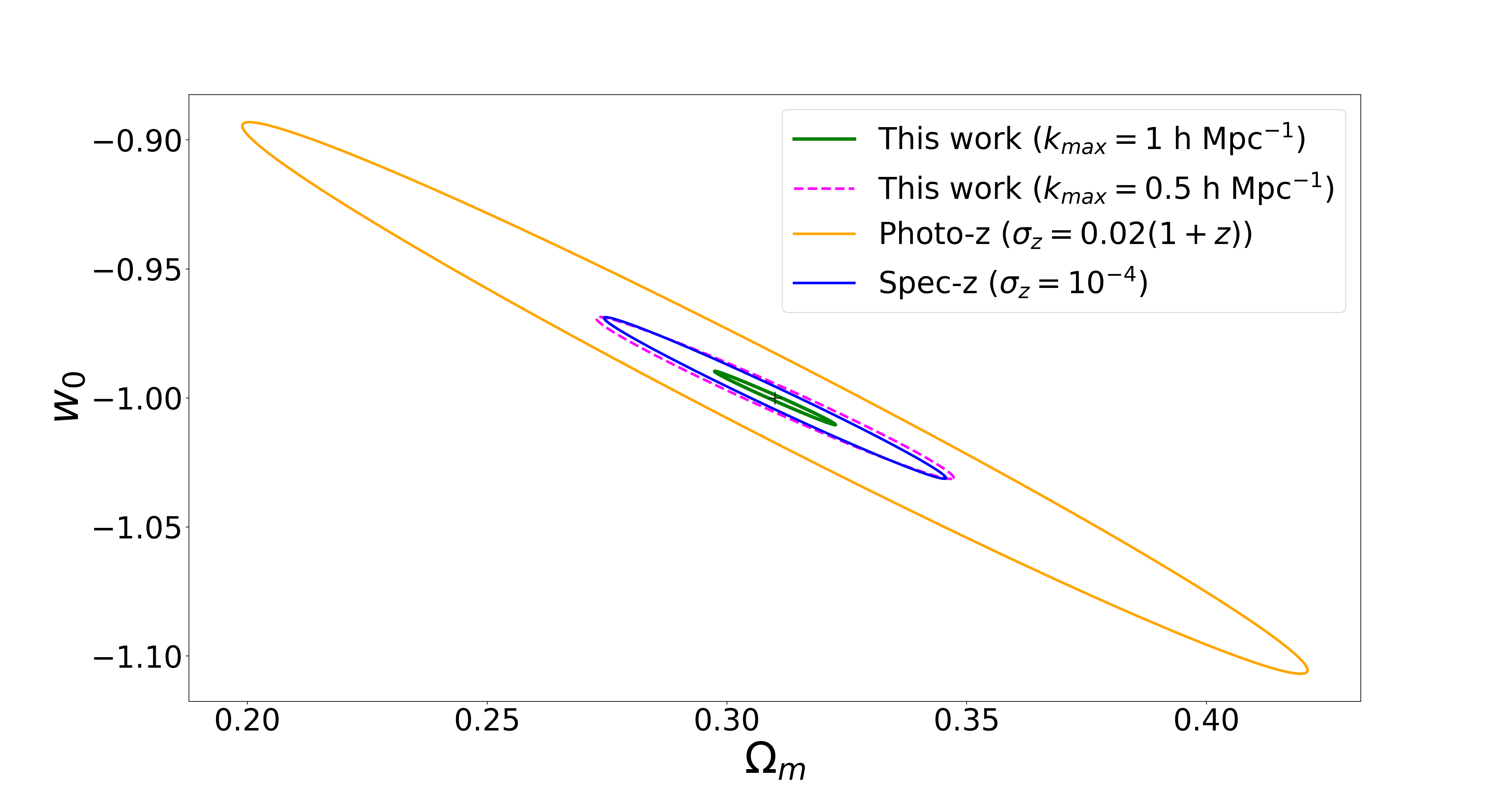}
     \end{subfigure}
  \begin{subfigure}{1.05\linewidth}
     \includegraphics[trim={0 0 0 0cm}, clip, width=1.\linewidth]{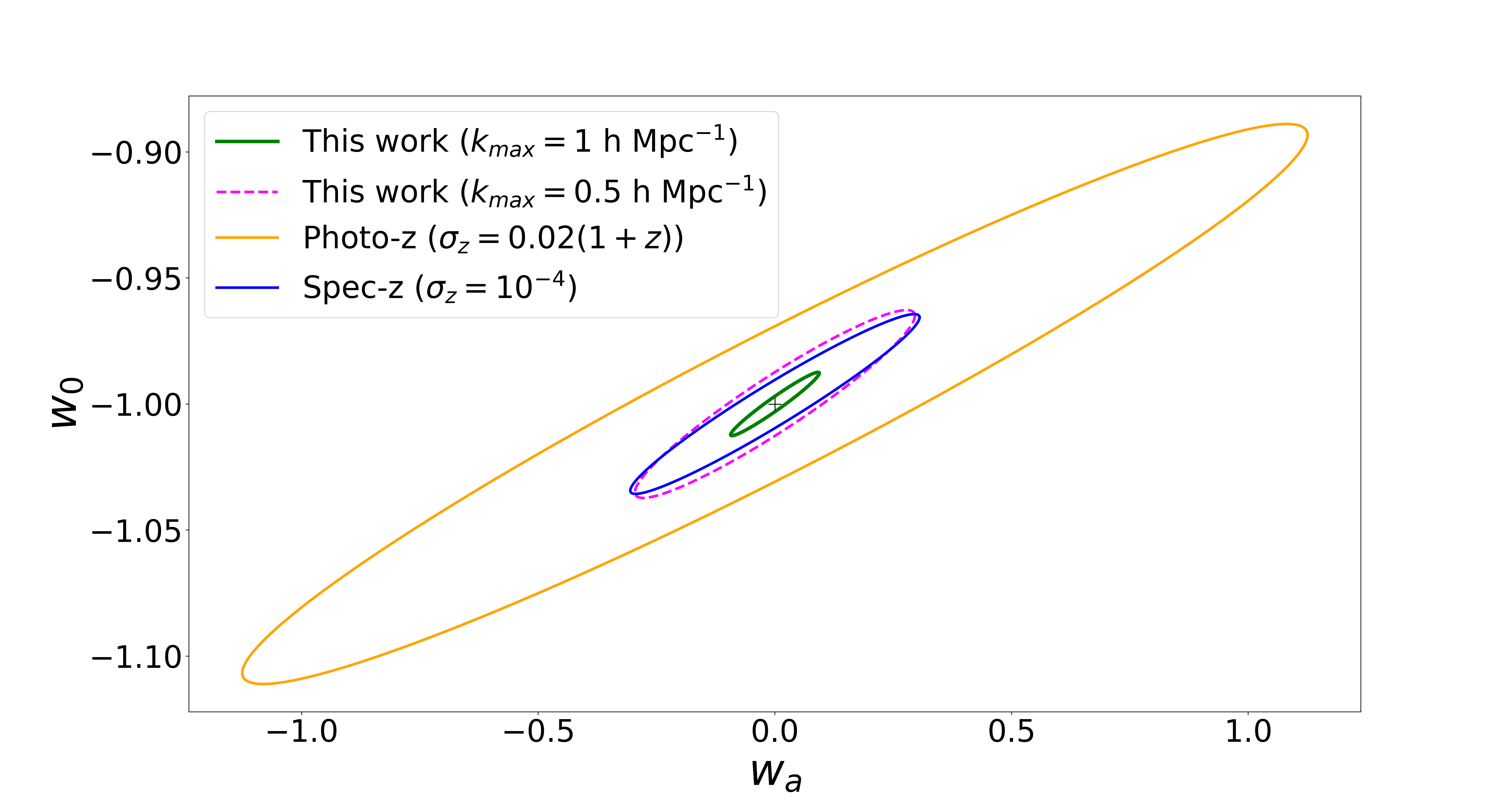}
     \end{subfigure}
     \captionsetup{singlelinecheck=on,justification=raggedright}
 \caption{We show the $68\%$ contour jointly over three cosmological parameters for the case with total {$5\times 10^4$} SN samples and {$20 \times 10^6$} galaxies for an optimistic and a pessimistic case.  SN and galaxy bias are both {$1.6$} and $\sigma^2_{int}=0.12$ mag. {For comparison we also show the standard $D_L$--$z$ forecasts with optimistic photometric and spectroscopic redshifts.}}\label{Fig:om-w-wa}
\end{figure}

Fig.~\ref{Fig:om-w-wa} shows the Fisher forecast $68\%$ error contours for  the cosmological parameters $\theta= \{\Omega_m, w_o, w_a\}$ for a fiducial LCDM cosmology with bias parameters $b_{g}= b_{s}= {1.6}$. We assumed {$5\times 10^4$} SNe over the  $z$ range 0.1 to 0.8. 

This indicates that $w_0$ can be constrained with an accuracy of better than $2.4\%$ and $w_a$ is constrained to an accuracy of better than $20\%$ by this method.  For comparison with the classical test with photometric  and spectroscopic SN redshifts, we perform a Fisher analysis for the same sets of cosmological parameters using the luminosity-distance relation  in Eq. \eqref{mag-dl}. The likelihood can be written in terms of the covariance matrix $\Sigma$ for number of SNe samples $N_{sn}$ as \cite{Betoule:2014frx, Scolnic:2017caz}
\begin{align}\label{likelihood-sn1}
\begin{split}
\mathcal{K}(\bm{\theta}) \propto & \exp\bigg(-\frac{1}{2}{(\bm{\hat m}- \bm{m})^{\dagger}\bm{\Sigma^{-1}}(\bm{\hat m} - \bm{m})\bigg)} , 
\end{split}
\end{align}
with the diagonal covariance matrix \cite{Betoule:2014frx, Scolnic:2017caz}
{\begin{align}\label{likelihood-sn1-cov}
\begin{split}
{\Sigma_{ij}}& = \bigg(\sigma_{int}^2 + \bigg(\frac{5\sigma_z}{z \log(10)}\bigg)^2\bigg) \delta_{ij}.
\end{split}
\end{align}} 
We  show in Fig. \ref{Fig:om-w-wa} that our method 
can perform significantly better than  the  $D_L$--$z$ test  with a very optimistic  photo-$z$ error ($\sigma_z= 0.02(1+z)$). Perhaps more unexpectedly, it even outperforms the $D_L$--$z$ test when all SN redshifts are spectroscopic. We discuss in the conclusions why this result follows naturally from the relationship of the classical  $D_L$--$z$ test to the generalization we propose in this paper.

\begin{figure}[t]
\centering
\begin{subfigure}{1.05\linewidth}
       \centering
       \includegraphics[trim={0 0 0 0cm}, clip, width=1.\linewidth,keepaspectratio=true]{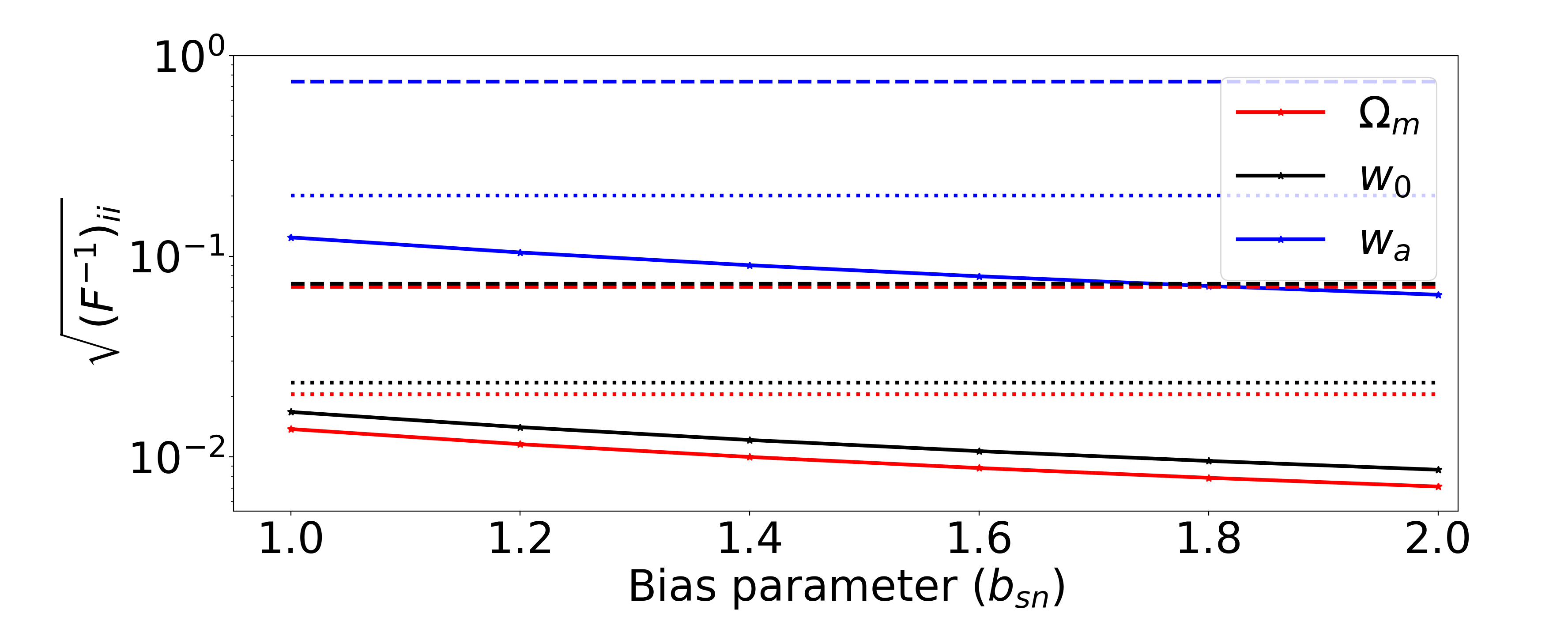}
     \caption{With $N_{sn}= 5\times 10^4$ and $N_{g}= 12\times 10^6$.}\label{Fig:om-w-wa-b-ns-1}
     \end{subfigure}
      \begin{subfigure}{1.05\linewidth}
      \centering
      \includegraphics[trim={0 0 0 0cm}, clip, width=1.\linewidth, keepaspectratio=true]{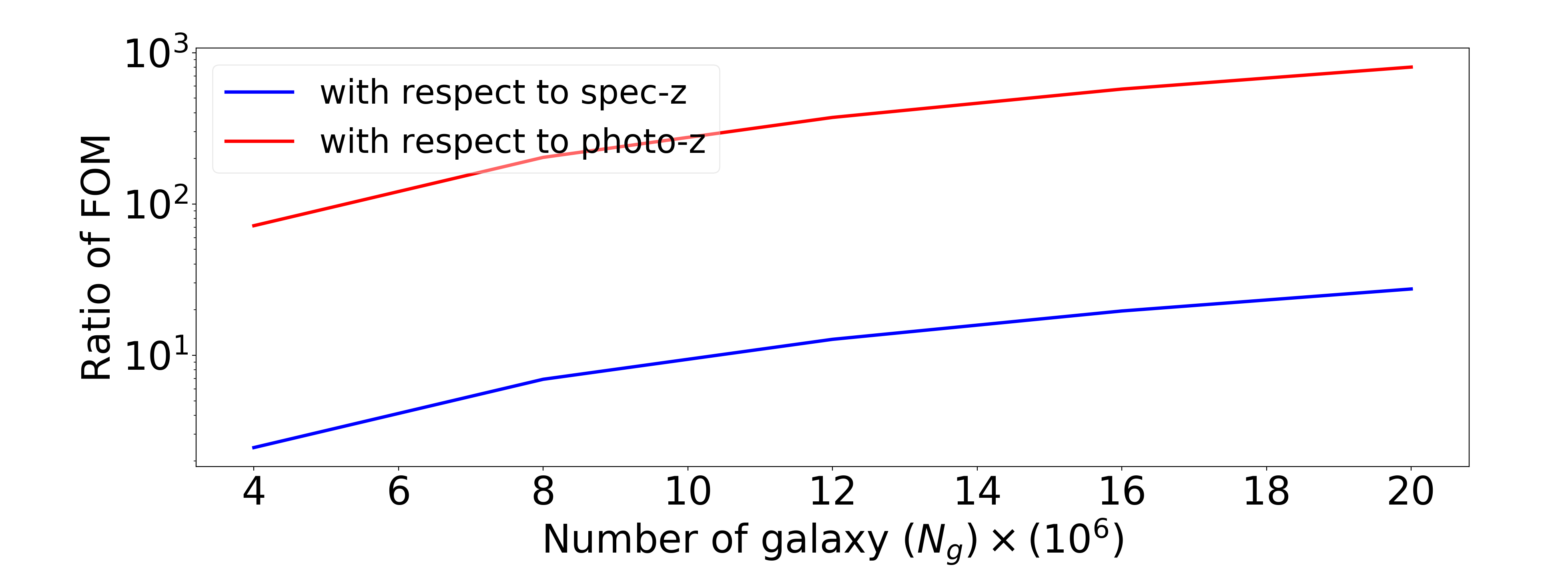}
     \caption{With $b_{\textrm{sn}} = b_{g}= 1.6$ and $N_{sn}=5\times 10^4$} \label{Fig:om-w-wa-b-ns-2} 
     \end{subfigure}   
\captionsetup{singlelinecheck=on,justification=raggedright}
\caption{(a) Change in Fisher forecast 
due to SN bias $b_{{s}}$ (solid) compared to  photometric (dotted) and  spectroscopic $D_L$--$z$ forecasts  (dashed).  (b) The ratio of Figure of Merit (FOM) of our method with the photometric 
and the spectroscopic 
case for different numbers of galaxies  $N_g$ sampling the volume of the SN survey. These results are  for $k_{max}= 1h$ Mpc$^{-1}$.}\label{Fig:om-w-wa-b-ns}
\end{figure}

We show the variation of our forecasts to SN bias in Fig. \ref{Fig:om-w-wa-b-ns-1}. Large bias implies stronger clustering and tightens the constraints. 

Another way to improve the constraint is to increase the number density of galaxy redshifts because the correlation function is larger for shorter pair distances.
The ratio of the Figure of Merit (FOM)= $(|\bm{F}|)^{1/2}$ between our method and the photometric or spectroscopic cases with varying $N_g$ are shown in Fig. \ref{Fig:om-w-wa-b-ns-2},  indicating that our method performs better than the photo-z and even the spec-z method for $k_{max}=1h$ Mpc$^{-1}$ for all $N_g>\,4\times 10^6$ when $b_{sn}\geq 1.6$.
 For larger numbers of SNe the relative errors between the different methods remain similar to the ones shown in Figs.~\ref{Fig:om-w-wa} and \ref{Fig:om-w-wa-b-ns}.

\paragraph{Discussion.} Photo-$z$ estimates for SNe are  vulnerable to several possible errors  due to dust extinction, intrinsic variability of the SNe spectrum, etc. As a result, a robust unbiased estimate of the true redshift of the object is difficult;  current methods produce a fraction of catastrophically wrong photo-$z$s. These contaminations are also difficult to simulate. Our proposed method is free from all these contaminations by construction. 

Our approach is also more robust to misidentified SNe. In the standard approach, SNe that are incorrectly identified as type Ia will tend to bias the  standard $D_L$--$z$ test. In our case, since SNe do not have redshift labels, they will be mapped to a far away place in comoving space where they will have no more than a random chance to find a galaxy to correlate to. The contamination arising from such wrongly typed SNe will therefore drop out of the cross-correlation on average and only add to the noise.

Our method affords the possibility of performing tests to probe for previously undiscovered systematics using the same data set. The properties of the SN may be correlated to galaxy properties like metallicity, color, etc, cf. \cite{2017arXiv170607697R}. Study of the correlation function with subsets of galaxies can be useful to infer whether there is any deviation in the clustering property of the SNe with different galaxies populations. The split of the data set will increase the error-bar of the estimate, but will be a new technique to find any environment dependence of the standard candles.
With future surveys like LSST any such effect can be probed  with higher statistical significance.

\paragraph{Conclusion.}\label{conclusion}
In this Letter, we propose a new method to probe the cosmological parameters with large supernova (SN) samples which have accurate estimates of the $D_L$s but no measurements of the source redshift, the regime of LSST. Our method relies on the idea that the SNe are tracers of the underlying matter distribution of the Universe and are therefore correlated with other tracers like galaxies. As a result, by measuring the parameter dependent cross-correlation of the SN and galaxy samples, we can statistically infer the cosmological parameters from the SNe. The stronger the spatial correlation between the SN and the galaxies, the better are the parameter estimates.

For the modest number of $5\times 10^4$ samples of SNe from LSST, and a galaxy  catalog with 4 million redshifts we can constrain dark energy parameters with an accuracy comparable with an optimistic forecast based on photo-$z$s.   \cite{Abell:2009aa} while being far more robust to photo-$z$ and type contamination systematics. With realistic values of the clustering bias, and for the galaxy redshift surveys we expect when LSST is operational (with 20 million galaxies) the cosmological parameters forecasts equal or exceed the performance of the standard $D_L$--$z$ test on a spectroscopic SN sample, see Fig. \ref{Fig:om-w-wa-b-ns}. As a result, this method is a powerful and robust way to infer cosmological parameters from large upcoming photometric SNe samples.

How is it possible that our cross-correlation A-P test outperforms  the classical spectroscopic $D_L$--$z$ test? Consider that our test weighs all pairs using a fiducial model of the correlation function; it reduces to the standard $D_L$--$z$ test in the limit of taking a fiducial model that ignores correlations between distance and redshift tracers except at zero distance. In that limit a galaxy is only counted if a SN went off in it, giving a unique combination of $D_L$ and $z$ at the same point, thus recovering the standard spectroscopic $D_L$--$z$ test that requires such pairs. In contrast, our generalization uses pairs at all distances for which there is non-zero correlation in the fiducial model; in addition, it exploits the isotropy of galaxy-SN pairs in co-moving space. 
We conclude that the spectroscopic $D_L$--$z$ test can be seen as  a cross-correlation A-P test with a single galaxy per SN, and an unphysical, sharply peaked fiducial correlation function with effective  $k_{\max}>10h$  Mpc$^{-1}$, in the deeply non-linear regime (scales shorter than the size of individual SN host galaxies). In spite of this, the classical $D_L$--$z$ test still works; we therefore expect the forecasts for our approach with modest $k_{\max}=1h$ Mpc$^{-1}$ to be very conservative. In case reliable redshift information is available for some subset of SN, this information can be included in our test to further improve constraints.

As a multi-tracer (in the form of Eq.~\eqref{eq:fullloglike}) or cross-correlation  generalization of the A-P test our method is applicable to samples of \textit{any} distance tracers with undetermined redshifts, provided they have small sky localization error. One such source class is binary black hole (BBHs) sirens; excellent distance indicators through their gravitational wave signal, but without electromagnetic counterpart to infer their redshift \cite{Holz:2005df, McWilliams:2011zs}. For this case, the error model, Eq. \eqref{errormodel-1}  must also include the error associated with the sky-localization, in order to correctly forecast the error bars of the cosmological parameters.
By cross-correlating the GW sources with galaxy redshift samples from the same sky patches, we can ultimately make use of the tens of thousands of BBHs that will be seen by LISA \cite{2017arXiv170200786A}, the  Einstein-Telescope \cite{et}, DECIGO \cite{sato2017status}, etc., to infer the cosmological parameters, thus avoiding the problem of host identification of the BBHs \cite{Petiteau:2011we}.

\begin{acknowledgments}
BDW wishes to acknowledge stimulating discussions with Robert Lupton and Saul Perlmutter. SM and BDW also acknowledge useful discussions with Rahul Biswas and David Spergel. The Flatiron Institute is supported by the Simons Foundation. This work is also supported by the Labex ILP (reference ANR-10-LABX-63) part of the Idex SUPER, and received financial state aid managed by the Agence Nationale de la Recherche, as part of the programme Investissements d'avenir under the reference ANR-11-IDEX-0004-02. 
\end{acknowledgments}

\def\urlprefix{}
\def\url#1{}
\bibliography{supernova_correlation}
\bibliographystyle{apsrev}
\end{document}